\documentclass[11pt,a4paper]{article}
\usepackage{jheppub}
\usepackage{amsmath}
\usepackage{multicol,bbm}
\usepackage{enumerate}
\usepackage{bibentry}

\graphicspath{{./Figs/}}

\def\be{\begin{equation}}
\def\ee{\end{equation}}
\def\ba{\begin{eqnarray}}
\def\ea{\end{eqnarray}}
\def\bea{\begin{eqnarray}}
\def\eea{\end{eqnarray}}
\def\eq{\begin{equation}}
\def\eqe{\end{equation}}
\def\eqa{\begin{eqnarray}}
\def\eqae{\end{eqnarray}}

\def\nl{\nonumber\\}

\def\l{\langle}
\def\r{\rangle}

\title{Loop Corrections to Soft Theorems in Gauge Theories and Gravity}
\author[a,b]{Song He}
\author[a,c]{ Yu-tin Huang}
\author[d]{Congkao Wen}
\affiliation[a]{School of Natural Sciences, Institute for Advanced
Study, Princeton, NJ 08540, USA} 
\affiliation[b]{Perimeter Institute for Theoretical Physics, Waterloo, ON N2L 2Y5,
Canada}
\affiliation[c]{Department of Physics and Astronomy, National Taiwan University, Taipei 10617, Taiwan, ROC} 
\affiliation[d]{Centre for Research in String Theory, Department of Physics, Queen Mary University of London, Mile End Road, London E1 4NS, UK}
\emailAdd{songhe@ias.edu,yutin@ias.edu,c.wen@qmul.ac.uk} 

\abstract{In this paper, we study loop corrections to the recently proposed new soft theorems of Cachazo and Strominger~\cite{CS}, for both gravity and gauge theory amplitudes. We first review the proof of its tree-level validity based on BCFW recursion relations, which also establishes an infinite series of universals soft functions for MHV amplitudes, and a generalization to supersymmetric cases.  For loop corrections, we focus on infrared finite, rational amplitudes at one loop, and apply recursion relations with boundary or double-pole contributions. For all-plus amplitudes, we prove that the subleading soft theorems are exact to all multiplicities for both gauge and gravity amplitudes. For single-minus amplitudes, while the subleading soft theorems are again exact for the minus-helicity soft leg, for plus-helicity loop corrections are required. Using recursion relations, we identify the source of such mismatch as stemming from the special contribution containing double poles, and obtain the all multiplicity one-loop corrections to the subleading soft behavior in Yang-Mills theory.  We also comment on the derivation of soft theorems using BCFW recursion in arbitrary dimensions.}
\preprint{QMUL-PH-14-12}

\begin{document}
\maketitle 
%%%%%%%%%%%%%%%%%%%%%%%%%%%%%%%%%%%%%%%%%%%%%%%%%%%%%%%%%%%%%
\section{Introduction and motivations}
%%%%%%%%%%%%%%%%%%%%%%%%%%%%%%%%%%%%%%%%%%%%%%%%%%%%%%%%%%%%%
It is well known that gauge and gravity scattering amplitudes diverge as one of the external momenta becomes soft. The leading divergences are well studied and  in general are universal~\cite{Weinberg, BernSoft, KosowerSoft}, meaning independent of the helicities of the remaining (hard) legs. These ``soft theorems" can be used in turn to constrain ansatz for unknown scattering amplitudes~\cite{SoftConstrain}, and in some cases, iteratively construct the full amplitude~\cite{SoftConstructable}. 

In general, soft behaviors are intimately tied to symmetry properties of the theory. A prominent example is the ``Adler's zero"~\cite{AdlerZero} which states that scalar amplitudes of non-linear sigma models must vanish as one of the scalar momentum is taken to be soft. This vanishing behavior is a reflection of the degeneracy of vacua. When the global symmetry is broken, due to anomalies, the soft limit no longer vanishes~\cite{RaduAnomaly}. For graviton scattering amplitude, the leading soft-singularity is governed by the ``Weinberg's theorem"~\cite{Weinberg}, which is stated as:\footnote{Here all gravity amplitudes (and similarly for Yang-Mills) are stripped of their coupling constant dependence, $\mathcal{M}_n=i \kappa^{n+2L-2} M_n $. We will also suppress factors of $1/(4\pi)^{2L}$ for loop-amplitudes.}
\eq\label{WienbergSoft}
M_{n+1}(1,\cdots, n,\epsilon s)=\frac{1}{\epsilon}S_{\rm G}^{(0)}M_n(1,\cdots,n)+\mathcal{O}(\epsilon^0)\,.
\eqe  
where we take leg $s$ to be soft, and parametrize this limit by $\epsilon\rightarrow 0$. The ``Soft function" $S_{\rm G}^{(0)}$ is defined as:
\eq
S_{\rm G}^{(0)}=\sum_{i=1}^n \frac{E_{s}^{\mu\nu} k_{i\mu}k_{i\nu}}{ s\cdot k_i}\,.
\eqe
The soft behavior in eq.(\ref{WienbergSoft}) has recently been understood as a consequence of Bondi, van der Burg, Metzner, and Sachs (BMS) symmetry \cite{BMS, Strominger1, Strominger2, Strominger3}, which is a large diffeomorphism transformations that translate between asymptotic flat solutions. It has been further argued that when the BMS transformations are taken as a whole, should imply an extension of the soft theorem to order $\epsilon^0$ with its coefficient given by the action of $S_{\rm G}^{(1)}$ acting on the $n$-point amplitude, where the operator $S_{\rm G}^{(1)}$ is given by~\cite{Vnote}
\eq
S_{\rm G}^{(1)}=\sum_{i=1}^n \frac{E_{s}^{\mu\nu} k_{i\mu}(s^\rho J_{i\nu\rho})}{ s\cdot k_i}\,.
\eqe
This sub-leading soft graviton theorem was recently discussed by White using eikonal methods~\cite{White:2011yy}, and its validity has been verified for all tree-level graviton amplitudes in~\cite{CS}.  Quite interestingly in the analysis of~\cite{CS} it was revealed that the series of universal soft functions can be further extended to order $\epsilon^1$:
\eq\label{BMSSoft}
M_{n+1}(1,\cdots, n,\epsilon s)=\frac{1}{\epsilon}S_{\rm G}^{(0)}M_n(1,\cdots,n)+S_{\rm G}^{(1)}M_n(1,\cdots,n)+\epsilon S_{\rm G}^{(2)}M_n(1,\cdots,n)+\mathcal{O}(\epsilon^2)\,.
\eqe
where
\eq
S_{\rm G}^{(2)}=\frac{1}{2}\sum_{i=1}^n \frac{E_{s}^{\mu\nu}(s^\rho J_{i\nu\rho})(s^\sigma J_{i\mu\sigma})}{ s\cdot k_i}\,.
\eqe
Interestingly, the leading order tree-level gluon soft function can be understood through similar symmetry arguments~\cite{Strominger3}. A recent analysis of soft behaviours for gluon tree-level amplitudes also revealed universal subleading soft-behaviors \cite{Casali:2014xpa}:
\eq\label{YMSoft}
A_{n+1}(1,\cdots, n,\epsilon s)=\frac{1}{\epsilon}S_{\rm YM}^{(0)}A_n(1,\cdots,n)+S_{\rm YM}^{(1)}A_n(1,\cdots,n)+\mathcal{O}(\epsilon^1)\,.
\eqe  
where 
\eq
S_{\rm YM}^{(0)}=\sum_{i=n, 1}\frac{E_{s}\cdot k_i}{s\cdot k_i},\quad S_{\rm YM}^{(1)}=\sum_{i=n, 1}\frac{E_{s\nu} (s_\mu J_i^{\mu\nu})}{s\cdot k_i}\,.
\eqe
Both gauge and gravity subleading soft theorems have been studied previously, e.g. in \cite{Low, Low2, SoftGravy1}. More recently they have been proven to hold in arbitrary dimensions~\cite{Schwab:2014xua}, based on scattering equations~\cite{CHY}. When written in terms of spinor helicity, the operator $S_{\rm YM}^{(1)}$ can be written in a bi-local form very similar to that of Yangian generators for $\mathcal{N}=4$ SYM~\cite{Henn}. Note that so far, both $S_{\rm G}^{(2)}$ and $S_{\rm YM}^{(1)}$ do not have a symmetry based understanding.

Since some of the above soft theorems are derived from symmetry arguments, an obvious question is whether it is an exact statement in perturbation theory. Naively, one would expect loop corrections to the subleading soft functions. Indeed for Yang-Mills, it is well known that even $S_{\rm YM}^{(0)}$ is already corrected at one loop. While for gravity, due to its dimensionful coupling, the dimensional analysis requires loop corrections to be suppressed by extra powers of soft-invariants relative to Yang-Mills~\cite{BernAllPlus}. Thus as a result, $S_{\rm G}^{(0)}$ is exact to all orders. However, the same analysis would imply that all subleading terms are corrected at one loop. 

To clarify the status of soft theorems at one loop, we consider rational terms in one-loop gauge and gravity amplitudes. In particular, to avoid complications due to infrared divergences we consider the all-plus and single-minus amplitudes in both gauge and gravity theories. Using BCFW-like recursion relations, which include boundary and double-pole contributions, we analytically prove that for both Yang-Mills and Gravity, the all-plus amplitudes satisfy the soft theorems exactly as written in eq.(\ref{BMSSoft}) and eq.(\ref{YMSoft}). For single-minus amplitudes, we find that if we take the the minus-helicity leg to be soft, the resulting soft behaviour again respects eq.(\ref{BMSSoft}) and eq.(\ref{YMSoft}). However, in the case where it is the positive-helicity leg that is taken to be soft, we find that for both gauge and gravity theories, eq.(\ref{BMSSoft}) and eq.(\ref{YMSoft}) no longer holds. From the recursive representation, it is clear that the source of such violation stems from the presence of double poles, which albeit do not have residues in real kinematics, do so in complex kinematics which is utilized for the recursion formula. We show that this is the only source of violation for the single-minus amplitudes, and derived suitable one-loop modification of the soft-functions.

This paper is organized as follows: In the next section we give a brief review of the tree-level soft theorems. Using the fact that the only non-trivial soft contributions in the BCFW representation stems from those with two-particle channels involving the soft leg, we extract the divergent terms by rewriting such contribution as an exponentiated operator acting on the $n$-point amplitude. The resulting expansion automatically reproduces eq.(\ref{BMSSoft}) and eq.(\ref{YMSoft}). A side result of this analysis is that for MHV amplitudes, where the BCFW recursion consists solely of such terms, there exists an infinite series of soft functions. We also discuss generalizations to supersymmetric amplitudes. In section~\ref{sec:softYM}, we move on to one-loop rational terms in pure gauge theory. We verify the validity of eq.(\ref{YMSoft}) for all plus amplitudes, while corrections is required for the single-minus amplitudes when the positive-helicity leg is taken to be soft. We derive the necessary one-loop correction terms for the single-minus amplitudes. In section~\ref{sec:softGrav} we repeat the same analysis with gravity rational terms, and reach the same conclusions. We close with conclusion and future directions in section~\ref{sec:Con}.

Added Note: In the completion of this manuscript, the work of Bern, Davies and Nohle~\cite{Bern:2014oka} appeared on the arXiv which has some overlap with the results here. In particular, while the validity of the soft theorem for all plus amplitudes was also verified numerically, the need for loop corrections were pointed out. Corrections related to the infrared divergence part of the amplitude was derived, and it was argued that such corrections are one-loop protected for $S_{\rm G}^{(1)}$ and two-loop protected for $S_{\rm G}^{(2)}$.

%%%%%%%%%%%%%%%%%%%%%%%%%%%%%%%%%%%%%%%%%%%%%%%%%%%%%%%%%%%%%
\section{Soft theorems for tree-level amplitudes}
%%%%%%%%%%%%%%%%%%%%%%%%%%%%%%%%%%%%%%%%%%%%%%%%%%%%%%%%%%%%%
%%%%%%%%%%%%%%%%%%%%%%%%%%%%%%%%%%%%%%%%%%%%%%%%%%%%%%%%%%%%%
\subsection{Review on Cachazo-Strominger Soft theorem}
%%%%%%%%%%%%%%%%%%%%%%%%%%%%%%%%%%%%%%%%%%%%%%%%%%%%%%%%%%%%
The Cachazo-Strominger tree-level soft theorem can be most easily derived via BCFW representation of tree amplitudes, where the singular part of the amplitude in the soft limit is given solely by two-particle factorization channels. Here we present a brief review. 

Consider the BCFW representation for tree-level gravity amplitude.\footnote{For the BCFW recursion relations in gravity amplitudes, see \cite{Bedford:2005yy, Benincasa:2007qj, ArkaniHamed:2008yf}.} Choosing the would be soft-graviton with positive-helicity to be shifted holomorphicaly,
\bea \label{BCFWshift}
{\lambda}_{\hat{s}} = 
{\lambda}_{s} +  z \lambda_{n}\,,\quad \tilde{\lambda}_{\hat{n}} = 
\tilde{\lambda}_{{n}} -  z \tilde{\lambda}_{s}  \, ,
\eea
the BCFW representation is given by:
\eq
\label{BCFWrep}
M_{n+1}(1,2,\ldots,n, s^+)=\sum_{1\leq i<n}M_3(\hat{s}^+,i,-\hat{K}_{i s})\frac{1}{K^2_{i, s}} M_{n}(\hat{K}_{i s},\ldots,\hat{n})+R\,,
\eqe
where $R$ represents terms arising from factorization poles $1/(k_s+K)^2$ where $K$ is a non-null momentum. The shift (\ref{BCFWshift}) indicates that $M_3$ has to be the anti-MHV amplitude, and we define $K_{i, j}\equiv k_i{+}k_j$. The holomorphic soft limit is achieved by scaling $\lambda_s\rightarrow \epsilon\lambda_s$, and using momentum conservation to solve two anti-holomorphic spinors $(\tilde{\lambda}_a,\tilde{\lambda}_b)$ in terms of other $\tilde{\lambda}$'s. Following~\cite{CS}, we will choose to express $\tilde{\lambda}_1$ and $\tilde{\lambda}_2$ as linear combinations of the remaining anti-holomorphic spinors:
\eq
\tilde{\lambda}_1=-\sum_{i=3}^n \frac{\l 2i\r}{\l 21\r}\tilde{\lambda}_i-\epsilon \frac{\l 2s\r}{\l 21\r}\tilde{\lambda}_s,\quad \tilde{\lambda}_2=-\sum_{i=3}^n \frac{\l 1i\r}{\l 12\r}\tilde{\lambda}_i-\epsilon \frac{\l 1s\r}{\l 12\r}\tilde{\lambda}_s\,.
\eqe
Thus the amplitudes $M_n$ on both side of eq.~(\ref{BCFWrep}) is to be understood as a function of $n$ holomorphic and only $n{-}2$ anti-holomorphic spinors.  
\begin{figure}
\begin{center}
\includegraphics[scale=0.4]{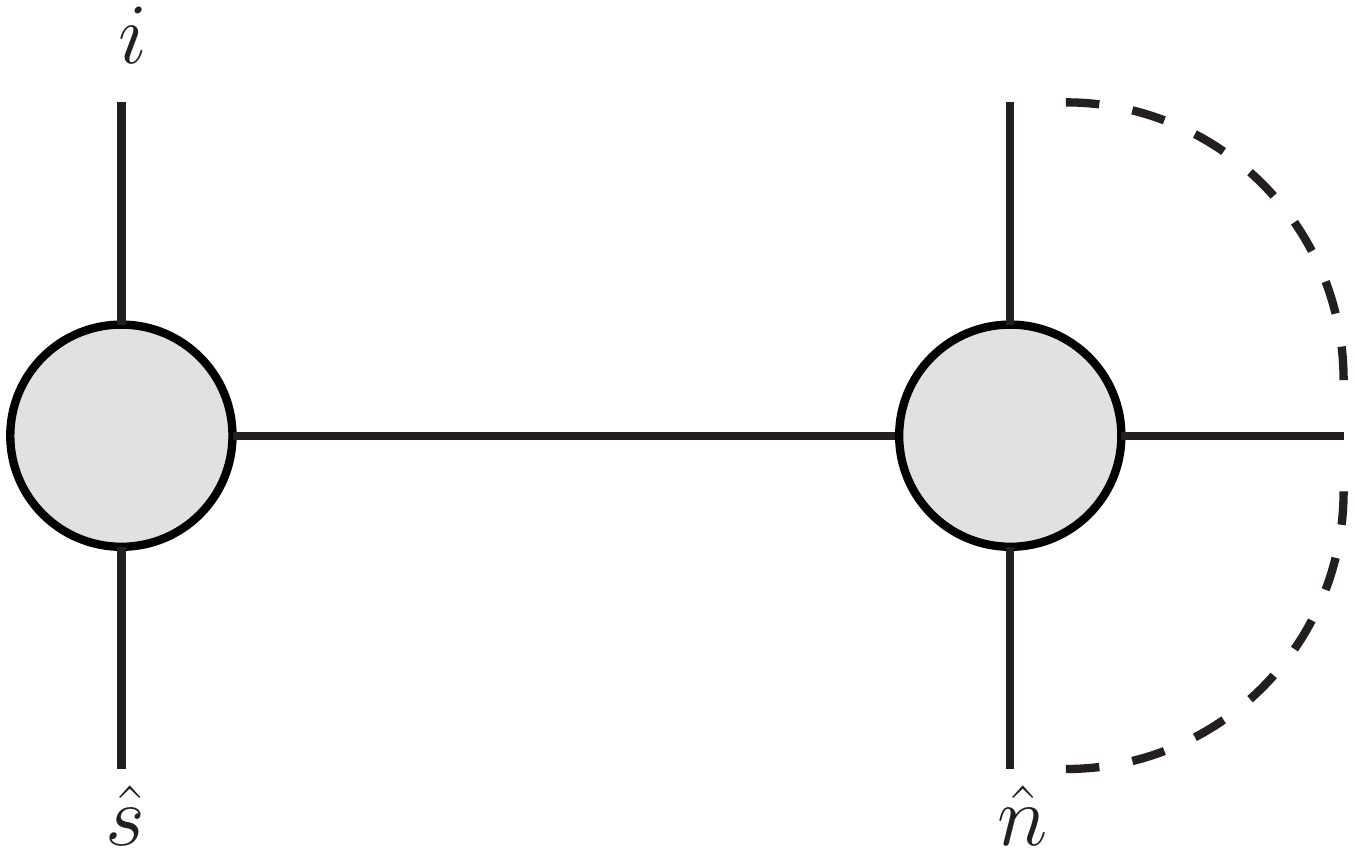}
\caption{\it
A particular factorization channel in the BCFW representation of a tree-level amplitude. }
\label{BCFWDia}
\end{center}
\end{figure}

It was shown explicitly in~\cite{CS} that the function $R$ is finite under the holomorphic soft limit, thus 
\eq\label{BCFWDiv}
M_{n+1}(1,2,\ldots,n,\{\epsilon\lambda_s,\tilde\lambda_s\}^+)\bigg|_{\rm div}=\sum_{1\leq i<n}M_3(\hat{s}^+,i,-\hat{K}_{i s})\frac{1}{K^2_{i, s}} M_{n}(\hat{K}_{i s},\cdots,\hat{n})\bigg|_{\rm div}
\eqe
where the shifted spinor ${\lambda}_{\hat{s}}$ is now defined as ${\lambda}_{\hat{s}} = 
\epsilon{\lambda}_{s} +  z \lambda_{n}$. For a given BCFW term in eq.(\ref{BCFWDiv}), represented in fig.\ref{BCFWDia}, the result can be written in a suggestive inverse soft form
\bea \nonumber
%M_{n+1}(\hat{s}^+ \, i | 3 \, \ldots \, \bar{n} ) 
M_3(\hat{s}^+,i, -\hat{K}_{i s} )\frac{1}{K^2_{is}}M_{n}( \hat{K}_{i s},\ldots,\hat{n})= \mathcal{S}_{s,i}
M_{n}( 
\{  \lambda_i, 
\tilde{\lambda}_i + { \langle s \, n \rangle \over \langle i \, n\rangle } \tilde{\lambda}_s \}   , 
\, \ldots \, , 
\{  \lambda_n, 
\tilde{\lambda}_n + { \langle s \, i \rangle \over \langle n \, i\rangle } \tilde{\lambda}_s \} ) \, , \\
\eea
where ``$\ldots$" part indicates un-shifted $\{ \lambda\, , \tilde{\lambda} \}$, and $\mathcal{S}_{s,i}$ is the ``inverse-soft-function" that is independent of the helicity of the $i$-th leg:\footnote{Note $\mathcal{S}_{s,i}$
may be brought into a different form by applying bonus relation of gravity amplitudes to the BCFW diagram Fig. \ref{BCFWDia} with $i=n{-}1$, which gives an extra factor, and leads to~\cite{Nandan:2012rk}
\bea
\mathcal{S}_{s,i} { \langle s \, n \rangle  \langle i \, n{-}1 \rangle \over \langle n i\rangle \langle s \, n{-}1\rangle  }
 = 
 { \langle n \, i \rangle \langle i \, n{-}1\rangle [ s\, i] \over \langle n\, s\rangle \langle s \, n{-}1\rangle  \langle s \, i\rangle  } \, .\nonumber
\eea }
\eq
\mathcal{S}_{s,i}=\frac{\langle ni\rangle^2 }{\langle ns\rangle^2}\frac{[i s]}{\langle i s\rangle}\;.
\eqe
First, note that the deformations on $\tilde{\lambda}_i$ and $\tilde{\lambda}_n$ can be written as an exponentiated differential operator acting on the unshifted amplitude:  
\bea
\nonumber &&M_{n}( 
\{  \lambda_i, 
\tilde{\lambda}_i + { \langle s \, n \rangle \over \langle i \, n\rangle } \tilde{\lambda}_s \}   , 
\, \ldots \, , 
\{  \lambda_n, 
\tilde{\lambda}_n + { \langle s \, i \rangle \over \langle n \, i\rangle } \tilde{\lambda}_s \} )
\\
\nonumber &=&\quad
e^{{ \langle s \, n \rangle \over \langle i \, n\rangle } \tilde{\lambda}_s 
\cdot { \partial \over  \partial \tilde{\lambda}_{i} } +{ \langle s \, i \rangle \over \langle n \, i\rangle } \tilde{\lambda}_s 
\cdot { \partial \over  \partial \tilde{\lambda}_{n} } }
M_{n}( 
\{  \lambda_i, 
\tilde{\lambda}_i  \}   , 
\, \ldots \, , 
\{  \lambda_n, \tilde\lambda_n \} ) \,.
\eea
Using this representation it is straight forward to obtain the divergent part of the holomorphic soft-limit
\eqa\label{CSsoft}
\nonumber M_{n+1}(1,\ldots,n^-,\{ \epsilon\lambda_s, \tilde{\lambda}_s\}^+)\bigg|_{\rm div}&=&\sum^{n-1}_{i=1}\frac{1}{\epsilon^3}\frac{\langle ni\rangle^2 }{\langle ns\rangle^2}\frac{[i,s]}{\langle i,s\rangle}e^{\epsilon\left({ \langle s \, n \rangle \over \langle i \, n\rangle } \tilde{\lambda}_s 
\cdot { \partial \over  \partial \tilde{\lambda}_{i} } +{ \langle s \, i \rangle \over \langle n \, i\rangle } \tilde{\lambda}_s 
\cdot { \partial \over  \partial \tilde{\lambda}_{n} } \right)}M_{n} \bigg|_{\rm div}\\
&=&\left(\frac{1}{\epsilon^3}S_{\rm G}^{(0)}+\frac{1}{\epsilon^2}S_{\rm G}^{(1)}+\frac{1}{\epsilon}S_{\rm G}^{(2)}\right)M_{n}
\eqae
where the operates $S_{\rm G}^{(k)}$ is defined as:
\bea
S_{\rm G}^{(k)} = 
\sum^{n-1}_{i=1} {1 \over k! }\mathcal{S}_{s,i}
\left({ \langle s \, n \rangle \over \langle i \, n\rangle } \tilde{\lambda}_s 
\cdot { \partial \over  \partial \tilde{\lambda}_{i} } +{ \langle s \, i \rangle \over \langle n \, i\rangle } \tilde{\lambda}_s 
\cdot { \partial \over  \partial \tilde{\lambda}_{n} } \right)^k  \, .
\eea
For $k=1$, we have
\eq
S_{\rm G}^{(1)}= 
\sum^{n-1}_{i=1} 
\left(\frac{\langle ni\rangle }{\langle ns\rangle}\frac{[i\,s]}{\langle i\,s\rangle} \tilde{\lambda}_s 
\cdot { \partial \over  \partial \tilde{\lambda}_{i} } -\frac{\langle n \, i\rangle}{\langle n\, s\rangle^2}[i\,s]\tilde{\lambda}_s 
\cdot { \partial \over  \partial \tilde{\lambda}_{n} } \right)=\sum^{n-1}_{i=1} 
\frac{\langle n\, i\rangle }{\langle ns\rangle}\frac{[i \, s]}{\langle i\, s\rangle} \tilde{\lambda}_s 
\cdot { \partial \over  \partial \tilde{\lambda}_{i} }\,,
\eqe
where the second term in the first inequality vanishes due momentum conservation. For $k=2$ we have 
\bea
\nonumber S_{\rm G}^{(2)} &=& 
\sum^{n-1}_{i=1} {1 \over 2 }\frac{\langle ni\rangle^2 }{\langle ns\rangle^2}\frac{[i\,s]}{\langle i\,s\rangle}
\left({ \langle s \, n \rangle \over \langle i \, n\rangle } \tilde{\lambda}_s 
\cdot { \partial \over  \partial \tilde{\lambda}_{i} } +{ \langle s \, i \rangle \over \langle n \, i\rangle } \tilde{\lambda}_s 
\cdot { \partial \over  \partial \tilde{\lambda}_{n} } \right)^2  \, \\
&=& 
\sum^{n}_{i=1} {1 \over 2 }{[ i\,s]\over \l i\,s\r}\left(\tilde\lambda_s \cdot\frac{\partial}{\partial \tilde\lambda_i}\right)^2\,,
\eea
where the last equivalence was obtained using momentum conservation, $\sum_i p^{\alpha\dot{\beta}}_i=0$,  and Lorentz invariance, $\sum_i m_{i}^{\dot{\alpha}\dot{\beta}}=0$. This establishes the Cachazo-Strominger soft theorem for tree-level gravity amplitudes. One can obtain a similar result by taking anti-holomorphic soft limit for a negative-helicity graviton, which leads to the parity conjugate of eq.(\ref{CSsoft}). Note that for MHV amplitudes, the equilvalence in eq.~(\ref{BCFWDiv}) holds without truncating to the divergent terms in the holomorphic soft limit. Therefore for MHV scattering amplitudes the soft theorem can be extended to arbitrary orders in $\epsilon$ with the inclusion of an infinite series of soft operators $S_{\rm G}^{(k)}$: 
\eqa
 M_{n+1}^{\rm MHV}(1,\ldots,n,s^+)&=&\left(\sum^{\infty}_{k=0}\frac{1}{\epsilon^{3-k}}S_{\rm G}^{(k)}\right)M^{\rm MHV}_{n}( 
\{  \lambda_i, 
\tilde{\lambda}_i  \}   , 
\, \ldots \, , 
\{  \lambda_n ,\tilde{\lambda}_n\} )\,.
\eqae

The same analysis can be applied to the scattering amplitudes in Yang-Mills theories. Color-ordered Yang-Mills amplitude has milder soft-divergence compared to gravity in the holomorphic soft limit, due to the milder little-group rescaling. A recent analysis~\cite{Casali:2014xpa} has shown that the subleading soft behavior of a $(n{+}1)$-point Yang-Mills tree amplitude can also be written as an operator acting on a $n$-point tree amplitude. 

The soft gluon theorem can be derived in a parallel fashion with gravity by using the BCFW representation of tree-level amplitudes, and the divergent term under the holomorphic soft limit is again isolated to the two particle channel indicated in Fig.~\ref{BCFWDia}(only one term $i=1$ contributes because of the color ordering):
\bea
\nonumber
A_{n{+}1}(1\,,\ldots\,,n,s^+))\bigg|_{\rm div}&=&
A_3(\hat{s}^+,1, -\hat{K}_{s1})\frac 1 {K^2_{1s}} A_n(\hat{K}_{s1},\ldots ,\hat{n} )\bigg|_{\rm div}\nl
&=&{ \langle n 1 \rangle  \over \epsilon^2\langle n s\rangle \langle s 1\rangle  }
A_{n}( 
\{  \lambda_1, 
\tilde{\lambda}_1 
 + \epsilon { \langle s \, n \rangle \over \langle 1 \, n\rangle } \tilde{\lambda}_s \}   , 
\ldots, 
\{  \lambda_n, \tilde{\lambda}_{n}
 +\epsilon { \langle s \, 1 \rangle \over \langle n \, 1\rangle } \tilde{\lambda}_s \} )\bigg|_{\rm div} \nl
 &=&
 { \langle n 1 \rangle  \over \epsilon^2\langle n s\rangle \langle s 1\rangle  }
 e^{ \epsilon\left( { \langle s  n \rangle \over \langle 1 n\rangle } \tilde{\lambda}_s \cdot { \partial \over \partial 
 \tilde{\lambda}_1 } +
 { \langle s  1 \rangle \over \langle n  1\rangle } \tilde{\lambda}_s \cdot { \partial \over \partial
 \tilde{\lambda}_n }   \right)}
A_{n}%( \{  \lambda_1, \tilde{\lambda}_1  \} \ldots , \{  \lambda_n, \tilde{\lambda}_{n} \} ) 
 \bigg|_{\rm div}\, . 
\eea
Then by the same argument as that of gravity amplitudes, general Yang-Mills amplitudes under the soft limit satisfy
\bea
\nonumber A_{n+1}(  \{  \lambda_1, \tilde{\lambda}_1 \}\, , \ldots \,,  
\{  \lambda_n, \tilde{\lambda}_n \}\, , \{ \epsilon  \lambda_s,  \tilde{\lambda}_s \}^+ )\bigg|_{\rm div}
= \sum_{k=0,1} \frac{1}{\epsilon^{2-k}}S_{\rm YM}^{(k)} 
A_{n}(  \{  \lambda_1, \tilde{\lambda}_1 \}\, , \ldots \,,  
\{  \lambda_n, \tilde{\lambda}_n \}) \nonumber
\eea
with
\bea \label{SkYM}
S^{(k)}_{\rm YM} = 
{1 \over k!} { \langle n 1 \rangle  \over \langle n s\rangle \langle s 1\rangle  } 
\left( { \langle s  n \rangle \over \langle 1 n\rangle } \tilde{\lambda}_s \cdot { \partial \over \partial
 \tilde{\lambda}_1 } +
 { \langle s  1 \rangle \over \langle n  1\rangle } \tilde{\lambda}_s \cdot { \partial \over \partial
 \tilde{\lambda}_n }   \right)^k \, .
\eea
So for the tree-level amplitudes in Yang-Mills theories, only $S^{(0)}_{\rm YM}$ and $S^{(1)}_{\rm YM}$ are universal. Interestingly, the operator for $S_{\rm YM}^{(1)}$ can be written in a bifocal form that is reminiscent of an Yangian generator:
\eq
S_{\rm YM}^{(1)}=\frac{p_s^{\alpha\dot{\alpha}}}{\langle n s\r \langle s 1\r  [1 n]}\frac{1}{2}\left[ p_{n\alpha}\,^{\dot{\beta}}(M_{1\dot{\beta}\dot{\alpha}}-\epsilon_{\dot{\beta}\dot{\alpha}}(D_1-h_1))- n\leftrightarrow 1\right]\,,
\eqe
where $M_{1\dot{\beta}\dot{\alpha}}$, $D_1$ and $h_1$ are the Lorentz, dilatation and helicity generator defined on leg $1$ respectively. Note that unlike the level-1 Yangian generator, which involves a summation of the above parenthesis over all distinct pairs of legs, this only involve the leg $1$ and $n$, thus in general does not vanish when act on the $n$-point amplitude.

Furthermore, we note that the soft factors of Yang-Mills theory are tightly related to the gravity soft factors. In fact, the gravity soft factors can be expressed as a double copy of that of Yang-Mills theory, 
\bea\label{doublecopy}
{1 \over \epsilon^3 }S^{(0)}_{\rm G} + {1 \over \epsilon^2 } S^{(1)}_{\rm G} + {1 \over \epsilon } S^{(2)}_{\rm G}
= \sum^{n}_{i=1} K^2_{s i} \left( {1 \over \epsilon^2 } S^{(0)}_{\rm YM}(i, s, n) + {1 \over 2\epsilon } S^{(1)}_{\rm YM}(i, s, n)  \right)^2 \, ,
\eea
where $K^2_{s, i} = \epsilon \langle s i \r [s i]$, and $S^{(k)}_{\rm YM}(i, s, n)$ is given in eq.(\ref{SkYM}) with leg $1$ replaced by a general leg $i$. Given the recent advance on gravity amplitudes as a double copy of Yang-Mills amplitudes \cite{Bern:2008qj}, it would be very interesting to see whether there is a deeper physical reason behind this double copy relation\footnote{BCJ and double copy in the context of soft limit were also discussed in~\cite{Oxburgh:2012zr}.}.

\subsection{Soft theorems for tree-level supersymmetric amplitudes}

Similarly one can study $\mathcal{N}=4$ SYM by applying supersymmetric BCFW recursion relation~  \cite{ArkaniHamed:2008gz, Brandhuber:2008pf}, where the leading divergent part is still controlled by the BCFW diagram Fig.\ref{BCFWDia} (again with $i=1$). After performing fermionic $\eta$ integration for the internal line, we find~\cite{Nandan:2012rk}
\be
{\langle n 1 \rangle \over \langle n s\rangle \langle s 1 \rangle} \mathcal{A}_{n}( 
\{{\lambda}_1, \tilde{\lambda}_1 + {\langle s n \rangle \over \langle 1 n \rangle}, 
\eta_1 + {\langle s n \rangle \over \langle 1 n \rangle} \eta_s  \} ,  \dots , 
\{{\lambda}_n, \tilde{\lambda}_n + {\langle s 1 \rangle \over \langle n 1 \rangle} \tilde{\lambda}_s, 
\eta_n + {\langle s 1 \rangle \over \langle n 1 \rangle} \eta_s \} ), 
\ee
which again can be conveniently expressed in an exponentiated form,  
\be
{\langle n 1 \rangle \over \langle n s\rangle \langle s 1 \rangle} 
e^{{\langle s n \rangle \over \langle 1 n \rangle} 
\left( \tilde{\lambda}_s \cdot  { \partial \over \partial \tilde{\lambda}_1  } 
+  \eta_s \cdot  { \partial \over \partial \eta_1  }   \right) 
+ {\langle s 1 \rangle \over \langle n 1 \rangle} 
\left(   \tilde{\lambda}_s \cdot  { \partial \over \partial \tilde{\lambda}_n  } 
+  \eta_s \cdot  { \partial \over \partial \eta_n  }  \right) } \mathcal{A}_{n}(  \{  \lambda_1, \tilde{\lambda}_1 \}\, , \ldots \,,  
\{  \lambda_n, \tilde{\lambda}_n \}) \, .
\ee
We then have the super-symmetrized version of soft theorem in Yang-Mills theory under the holomorphic soft limit,
\bea
\nonumber \mathcal{A}_{n+1}(  \{  \lambda_1, \tilde{\lambda}_1 \}\, , \ldots \,,  
\{  \lambda_n, \tilde{\lambda}_n \}\, , \{ \epsilon  \lambda_s,  \tilde{\lambda}_s \} )\bigg|_{\rm div}
= \sum_{k=0,1} \frac{1}{\epsilon^{2-k}} \mathcal{S}_{\rm YM}^{(k)} 
\mathcal{A}_{n}(  \{  \lambda_1, \tilde{\lambda}_1 \}\, , \ldots \,,  
\{  \lambda_n, \tilde{\lambda}_n \}) \nonumber
\eea
with
\bea \label{SkYM}
\mathcal{S}^{(k)}_{\rm YM} = 
{1 \over k!} { \langle n 1 \rangle  \over \langle n s\rangle \langle s 1\rangle  } 
\left[ { \langle s  n \rangle \over \langle 1 n\rangle } \left( \tilde{\lambda}_s \cdot { \partial \over \partial
 \tilde{\lambda}_1 } + \eta_s \cdot  { \partial \over \partial \eta_1  }   \right)+
 { \langle s  1 \rangle \over \langle n  1\rangle } 
 \left( \tilde{\lambda}_s \cdot { \partial \over \partial
 \tilde{\lambda}_n } + \eta_s \cdot  { \partial \over \partial \eta_n  }   \right) \right]^k \, .
\eea 
For the soft positive gluon, we set $\eta_s =0$ it reduces to what we had previously, whereas for other species of particles the amplitudes have milder soft behaviour. 

A similar calculation for the parity flipped version of the previous case leads to anti-holomorphic soft theorem, which is given as
\bea
\nonumber \mathcal{A}_{n+1}(  \{  \lambda_1, \tilde{\lambda}_1 \}\, , \ldots \,,  
\{  \lambda_n, \tilde{\lambda}_n \}\, , \{   \lambda_s, \epsilon \tilde{\lambda}_s \} )\bigg|_{\rm div}
= \sum_{k=0,1} \frac{1}{\epsilon^{2-k}} \bar{\mathcal{S}}_{\rm YM}^{(k)} 
\mathcal{A}_{n}(  \{  \lambda_1, \tilde{\lambda}_1 \}\, , \ldots \,,  
\{  \lambda_n, \tilde{\lambda}_n \}) \nonumber
\eea
with
\bea \label{SkYM}
\bar{\mathcal{S}}^{(k)}_{\rm YM} = 
{1 \over k!} 
{[n 1] \over [n s][s 1]}  \delta^4(\eta_s + \epsilon {[n s] \over [1 n]} \eta_1 + \epsilon {[s 1] \over [1 n]} \eta_{n})
\left( { [ s  n ] \over [ 1 n ] } {\lambda}_s \cdot { \partial \over \partial
 {\lambda}_1 } +
 { [ s  1 ] \over [ n  1 ] } 
 {\lambda}_s \cdot { \partial \over \partial
 {\lambda}_n } \right)^k \, .
\eea 
For the soft negative-helicity gluon, we need pick $\eta^4_s$, so the above formula reduces to the non-supersymmetric version we had earlier. Again other species of particles have milder singularity because of the appearance of $\epsilon$ in the fermionic delta-function.

Now $\mathcal{N}=8$ super gravity can be studied in the same way, and we find, 
\bea
\nonumber \mathcal{M}_{n+1}(  \{  \lambda_1, \tilde{\lambda}_1 \}\, , \ldots \,,  
\{  \lambda_n, \tilde{\lambda}_n \}\, , \{ \epsilon  \lambda_s,  \tilde{\lambda}_s \} )\bigg|_{\rm div}
= \sum^2_{k=0} \frac{1}{\epsilon^{3-k}} \mathcal{S}_{\rm G}^{(k)} 
\mathcal{M}_{n}(  \{  \lambda_1, \tilde{\lambda}_1 \}\, , \ldots \,,  
\{  \lambda_n, \tilde{\lambda}_n \}) \nonumber
\eea
for holomorphic soft limit, and 
\bea
\nonumber \mathcal{M}_{n+1}(  \{  \lambda_1, \tilde{\lambda}_1 \}\, , \ldots \,,  
\{  \lambda_n, \tilde{\lambda}_n \}\, , \{   \lambda_s, \epsilon \tilde{\lambda}_s \} )\bigg|_{\rm div}
= \sum^2_{k=0} \frac{1}{\epsilon^{3-k}} \bar{\mathcal{S}}_{\rm G}^{(k)} 
\mathcal{M}_{n}(  \{  \lambda_1, \tilde{\lambda}_1 \}\, , \ldots \,,  
\{  \lambda_n, \tilde{\lambda}_n \}) \nonumber
\eea
for anti-holomorphic soft limit, where the soft operators $\mathcal{S}_{\rm G}^{(k)} $ as well as $\bar{\mathcal{S}}_{\rm G}^{(k)} $ can be expressed as a double copy of those of $\mathcal{N}=4$ SYM, exactly in the same form of eq.~(\ref{doublecopy}).

%%%%%%%%%%%%%%%%%%%%%%%%%%%%%%%%%%%%%%%%%%%%%%%%%%%%%%%%%%%%%%%%%%%%%%%%%%
\section{Soft theorem for one-loop rational amplitudes in Yang-Mills \label{sec:softYM}}
%%%%%%%%%%%%%%%%%%%%%%%%%%%%%%%%%%%%%%%%%%%%%%%%%%%%%%%%%%%%%%%%%%%%%%%

In this section, we study the soft gluon theorem for the two classes of infrared finite one-loop amplitudes in pure Yang-Mills theory: the all-plus as well as single-minus amplitudes. 
\subsection{All-plus amplitudes}
First we consider the soft gluon theorem for one-loop all-plus amplitudes in pure Yang-Mills theory. The color-ordered, all-plus amplitude can be put in a very compact form \cite{Bern:1993qk, Mahlon:1993si},\footnote{From now on we will use $A^{(1)}$ ($A^{(0)}$) to denote one-loop (tree-level) Yang-Mills amplitude, and similarly $M^{(1)}, M^{(0)}$ for gravity amplitudes. }
\be\label{allplus}
A^{(1)}_n(1^+\,,\ldots\,,n^+)=\frac 1 {3}\frac{h_n}{\l 1 2\r\cdots\l n 1\r}, \quad h_n\equiv \sum_{1\leq i<j<k<l\leq n} {\rm tr}_-(i j k l),
\ee
where we have ${\rm tr}_-(i j k l)=\l i j\r[j k]\l k l\r[l i]$ (similarly for its parity-conjugate ${\rm tr}_+)$. It is straightforward to prove directly that the soft theorem, valid at tree level, receives no quantum corrections for all-plus amplitude at one loop:
\be\label{soft_allplus}
A^{(1)}_{n{+}1}(1^+\,,\ldots\,, n^+,\{\epsilon \lambda_s, \tilde\lambda_{s}\}^+)\bigg|_{\rm div}=
\left(\frac{1}{\epsilon^2}S_{\rm YM}^{(0)} +\frac{1}{\epsilon} S_{\rm YM}^{(1)} \right) A^{(1)}_n (1^+\,,\ldots\,, n^+)\,.
\ee
The leading soft behavior is trivial. For the subleading soft term, note that $S_{\rm YM}^{(1)}$ in eq.(\ref{soft_allplus}) acts only on $\tilde\lambda_n$ and $\tilde\lambda_1$ respectively. Since $h_n$ is cyclic symmetric, one can shift its summation range to $2\leq i<j<k<l\leq 1$, such that when both $n$ and $1$ are present in the trace, and it is proportional to $\langle n1\rangle$ instead of $[n1]$. The result is
\be
S_{\rm YM}^{(1)}\, h_n=\sum_{1<i<j<k<n} \l i j\r[j k][s i] \left(\frac{\l k n\r}{\l n s\r}+\frac{\l k 1\r}{\l s 1\r}\right)= S_{\rm YM}^{(0)} \sum_{1<i<j<k<n} {\rm tr}_-(i j k s)%\l i j\r[j k]\l k s\r[s\,i],
\ee
where we have used Schouten identity. This, multiplied by the $n$-pt Parke-Taylor formula, precisely agrees with the subleading term on the LHS of (\ref{soft_allplus}), which receives contribution from all possible trace factors with the last entry being $s$, and is given precisely by:
\eq
A^{(1)}_{n{+}1}(1^+\,,\ldots\,,n^+,\{\epsilon \lambda_s, \tilde\lambda_{s}\}^+)\bigg|_{\epsilon^{-1}}=\frac 1 {\epsilon} \frac{S^{(0)}_{\rm YM}}{\prod_{i=1}^n\l i\, i+1\r} \sum_{1<i<j<k<n} {\rm tr}_-(i j k s)\,.
\eqe

Although in this case it is essentially trivial to prove the theorem, we would like to revisit the proof using properties of its recursion formula, which would be very useful for more involved cases, including single-minus Yang-Mills amplitudes as well as those of gravity amplitudes, as we will consider in the following sections.

In \cite{BernAllPlus}, the recursions for all-plus amplitudes have been phrased using non-adjacent BCFW shifts, but we find the one with adjacent shift, eq.(\ref{BCFWshift}), is most suitable for our purpose. Unlike tree-level amplitudes, all-plus amplitudes have non-vanishing residues at $z \rightarrow \infty$, which has been observed already in \cite{BernAllPlus} with non-adjacent BCFW shifts . Using the formula (\ref{allplus}), we find the contribution from pole at infinity gives,\footnote{The denominator of (\ref{allplus}) is linear in $z$ with $z\to \infty$. In the numerator, the dominant term is also linear in $z$, which come from terms with $k=n$ and $l=n{+}1$. } 
\be
\mathcal{B}^{\rm YM}_n=-\frac  1 3\,\frac 1 {\l 12\r\cdots \l n 1\r} \sum_{1<i<j\leq n} \l i j\r[j s][i s]\, .
\ee
Thus we have a recursion for (\ref{allplus}) with the homogeneous term and a boundary term $\mathcal{B}^{\rm YM}_n$,
\eq\label{AllPlusRecYM}
A^{(1)}_{n+1}(1^+\,,\ldots\,,n^+,s^+)= A_3^{(0)}(\hat{s}^+, 1^+,-\hat{K}_{s 1}^-)\frac{1}{K^2_{s,1}}A^{(1)}_{n}(\hat{K}_{s 1}^+\,,\ldots\,,\hat{n}^+)+\mathcal{B}^{\rm YM}_n\,.
\eqe 
 Note that $\mathcal{B}^{\rm YM}_n$ is obviously finite as we take the soft limit $\lambda_{n{+}1}\to 0$. By the same argument as that of tree-level amplitudes, we see that (\ref{soft_allplus}) holds, which is a direct consequence of the finiteness of the boundary term under the soft limit. 
%%%%%%%%%%%%%%%%%%%%%%%%%%%%%%%%%%%%%%%%%%%%%%%%%%%%%%%%%%%%%%%%%%%%%%%%%%
\subsection{Single-minus amplitudes}
%%%%%%%%%%%%%%%%%%%%%%%%%%%%%%%%%%%%%%%%%%%%%%%%%%%%%%%%%%%%%%%%%%%%%%%%%%

We proceed to study the soft gluon theorem for the one-loop single-minus amplitude, $A_{n{+}1}(1^-, 2^+,\cdots,(n{+}1)^+)$. There are two types of soft limit one can consider: the holomorphic one with, say $\lambda_{n{+}1}=\epsilon \lambda_s$ and the anti-holomorphic one with $\tilde\lambda_1=\epsilon\tilde\lambda_s$.\footnote{The anti-holomorphic soft limit is simply the parity conjugate of the holomorphic one for single-plus amplitude, $A_{n{+}1}(1^+, 2^-,\ldots,(n{+}1)^-)$.} 

Our strategy will be similar to the case of all-plus amplitudes: we will write down recursion relations and extract terms with soft divergences in both cases. We consider the second case first, and we recall that the leading soft divergence is given by the parity-conjugate of $S_{\rm YM}^{(0)}$ above, times an all-plus amplitude,
\be\label{leading_oneminus}
A^{(1)}_{n{+}1}(\{\lambda_1,\tilde\lambda_1=\epsilon \tilde\lambda_s\}^-, 2^+,\ldots,(n{+}1)^+)\bigg|_{\epsilon^{-2}}
=\frac{1}{\epsilon^2} \bar S_{\rm YM}^{(0)} A^{(1)}_n (2^+,\ldots, (n{+}1)^+),
\ee
where $\bar S_{\rm YM}^{(0)}=\bar S_{\rm YM}^{(0)}(n{+}1,s,2)\equiv \frac{[n{+}1\,2]}{[n{+}1\, s][s 2]}$. Eq.(\ref{leading_oneminus}) can be easily checked by using the explicit formula in \cite{Bern:2005ji}, or by recursions as we will see shortly. 

The main use of recursions is to study the subleading contributions.  To study the anti-holomorphic soft limit of $1$, it is natural to consider the following shift,
\be\label{shift2}
 \lambda_{\widehat{n{+}1}}=\lambda_{n{+}1}+z\lambda_1, \quad\tilde\lambda_{\hat 1}=\tilde\lambda_1-z \tilde\lambda_{n{+}1}\, .\ee

It is a known fact that, for a shift such as (\ref{shift2}), there is actually no contributions from the pole at infinity for the single-minus amplitudes \cite{Bern1Minus} (which we have explicitly checked by using the all-$n$ formula for single-minus amplitudes in \cite{Bern:2005ji} ), but rather it generates double poles. The BCFW recursion then gives factorization terms (including the homogenous term) and an additional contribution involving double poles:
\ba\label{rec_oneminus}
A^{(1)}_{n{+}1}&=&\phantom{+}
A^{(1)}_n (\hat{1}^-,\ldots,(n{-}1)^+, \hat{K}_{n,n{+}1}^+)\frac 1 {K^2_{n,n{+}1}}  A_3^{(0)}(-\hat{K}_{n,n{+}1}^-,n^+,(\widehat{n{+}1})^+)\nl
&&+ A^{(0)}_{3} (\hat{1}^-,2^+,-\hat{k}^-_{1,2}) \frac 1 {K^2_{1,2}}A^{(1)}_{n{-}2} (\hat{K}^+_{1,2},3^+,\ldots,(\widehat{n{+}1})^+)\nl
&&+\sum_{j=3}^{n{-}2} A^{(0)}_{j{+}1} (\hat{1}^-,\ldots,j^+,-\hat{K}^-_{1,\cdots,j}) \frac 1 {K^2_{1,\ldots,j}} A^{(1)}_{n{-}j} (\hat{K}^+_{1,\ldots,j},(j{+}1)^+,\ldots,(\widehat{n{+}1})^+)\nl
&&+A^{(0)}_n(\hat{1}^-,\ldots, \hat{K}^-_{n,n{+}1}) \frac1 {(K^2_{n,n{+}1})^2}  V_3^{(1)}(-\hat{K}^+_{n,n{+}1},n^+,(\widehat{n{+}1})^+) \left(1{+} K^2_{n,n{+}1} S_{\rm YM}^{(0)} \bar S_{\rm YM}^{(0)} \right),\nl
\ea 
where we have separated out BCFW channel of $j=2$ on purpose, which is given by the second line, and the last line corresponds to the special contribution with both simple and double poles; $V_3^{(1)}(1,2,3)\equiv \frac 1 {3 } [1,2][2,3][3,1]$ is the all-plus vertex, and the simple-pole contribution is dressed with two soft factors, $S_{\rm YM}^{(0)}(\hat{1},\hat{K}_{n,n{+}1},n{-}1)$ and $\bar S_{\rm YM}^{(0)}(n,-\hat{K}_{n,n{+}1},\widehat{n{+}1})$. 

The key point is that the second line on the RHS of eq.~(\ref{rec_oneminus}) is the only term that diverges in the limit $\tilde\lambda_1\to 0$, whereas others remain finite in this limit. The proof for this fact of the first and third lines follows directly the same argument as given in the Appendix of~\cite{CS}, except this is the parity-conjugate case with anti-holomorphic soft limit. 

As for the term on the last line, actually it manifestly has no soft divergence at $\tilde\lambda_1\to 0$. Thus we conclude under the anti-holomorphic, single-minus amplitude at one loop takes the same form as that of tree-level amplitudes,  
\be
A^{(1)}_{n{+}1}(\{\lambda_1, \epsilon \tilde\lambda_s\}^-, 2^+,\cdots,(n{+}1)^+)\bigg|_{\rm div}=
\left( \frac{1}{\epsilon^2}  \bar S_{\rm YM}^{(0)} {+} \frac{1}{\epsilon} \bar S_{\rm YM}^{(1)} \right) A^{(1)}_n (2^+,\ldots, (n{+}1)^+)\,,%+\mathcal{O}( \epsilon^0 ),
\ee
where 
\be 
\bar S_{\rm YM}^{(1)}=\bar S_{\rm YM}^{(1)}(n{+}1, s, 2)\equiv \frac 1 {[n{+}1\, s]}\lambda_s\cdot \frac{\partial}{\partial \lambda_{n{+}1} }{+}\frac 1 {[s 2]}\lambda_s\cdot \frac{\partial}{\partial \lambda_2 } \, .
\ee

Now we turn to the study of the holomorphic soft limit of $n{+}1$, and the suitable recursion for this is also given by (\ref{rec_oneminus}). By now we have become familiar with the fact that the homogeneous term on the first line of  (\ref{rec_oneminus}) gives the expected leading and subleading terms in the soft limit $\lambda_{n{+}1}\to 0$, if there is no soft divergence from other contributions. This is indeed the case for the second and third lines. 

However, for the last line of (\ref{rec_oneminus}), we find both simple-pole and double-pole contributions diverge in the soft limit $\lambda_{n{+}1}=\epsilon \lambda_s\to 0$, because
\be 
K_{n,n{+}1}^2=\epsilon [n\,n{+}1]\l n s\r,\quad V_3^{(1)}=\epsilon\frac{\l 1 s\r}{\l 1 n \r}[n\, n{+}1]^3,\quad \bar S^{(0)}_{\rm YM}=\frac{\l 1 n\r}{\epsilon \l 1 s\r[ n\,n{+}1]},
\ee
and all other factors remain finite. 

Thus, unlike the all-plus or the anti-holomorphic case, the holomorphic soft limit of single-minus amplitude receives corrections at subleading order, 
\be\label{YMoneminus_soft}
A^{(1)}_{n{+}1}(1^-,\cdots,n^+,\{\epsilon \lambda_s, \tilde\lambda_{n{+}1}\}^+)\bigg|_{\rm div}=\left(\frac{1}{\epsilon^2}S_{\rm YM}^{(0)}+\frac{1}{\epsilon}S_{\rm YM}^{(1)}\right) A^{(1)}_n (1^-,\cdots,n^+)+\Delta_n\,,%+\mathcal O(\epsilon^0),
\ee
where $\Delta_n$ is the $\mathcal O (\epsilon^{-1})$ contribution on the last line of (\ref{rec_oneminus}):
\ba\label{corr}
\Delta_n&=&-A^{(0)}_n (1^-,2^+, \cdots,(n{-}1)^+, n^-)\times \bar{S}^{(0)}_{\rm YM} (n{-}1, n, n{+}1)\times \frac{[n\,n{+}1]}{\l n\, n{+}1 \r}\nl 
&=&-{1 \over \epsilon} \frac{\l n 1 \r^3 \l n{-}1\,s\r [n\, n{+}1]}{\l 1 2\r\ldots\l n{-}2\, n{-}1\r\l n{-}1 n\r^2 \l n s\r^2} \, .
\ea
For $n=4$, our general formula agrees with a special case computed in \cite{Bern:2014oka}.
The appearance of such correction to the subleading divergence is expected: as we have identified the source of the correction is due to the non-trivial poles (include double poles) of single-minus loop amplitudes, which is typical for non-supersymmetric gauge theory at loop level, but would not be present for all-plus amplitudes. It would be interesting to interpret eq.~(\ref{corr}) as coming from a correction of the operator $S^{(1)}_{\rm YM}$ on the RHS of (\ref{YMoneminus_soft}).

\section{Soft theorem for one-loop rational amplitudes in Gravity \label{sec:softGrav}}
In this section we will study one-loop all-plus and single-minus amplitudes in gravity theory. We will follow the same strategy for Yang-Mills theory discussed in the previous section: we will write down a BCFW-like recursion relation for the amplitudes, then carefully exam the soft-limit behaviour for each term in the recursion relation. 
 %%%%%%%%%%%%%%%%%%%%%%%%%%%%%%%%%%%%%%%%%%%%%%%%%%%%%%%%%%%%%%%%%%%%%%%%%%%%
 \subsection{All-plus amplitudes}
 %%%%%%%%%%%%%%%%%%%%%%%%%%%%%%%%%%%%%%%%%%%%%%%%%%%%%%%%%%%%%%%%%%%%%%%%%%%%
The $n$ positive-helicity graviton amplitude at one-loop is given by~\cite{BernAllPlus},
\be\label{gra_allplus}
M^{(1)}_n(1^+,\cdots,n^+)=\frac{(-)^{n{+}1}}{960}\sum_{a,M,b,N} H(a,\{M\},b)\,H(b,\{N\},a)\,{\rm tr}^3(a, \{M\}, b, \{N\}),
\ee
where the sum is over $1\leq a<b\leq n$ and over subsets $M,N$ satisfying $M\cap N=\emptyset$ and $M\cup N=\{1,\ldots,a{-}1,a{+}1,\ldots,b{-}1, b{+}1,\ldots,n\}$.  The function $H(a,\{M\},b)$ will be discussed below, and the trace is defined as
\be {\rm tr} (a, \{M\}, b,\{ N\})=\sum_{i\in M,\, j\in N} {\rm tr} (a i b j), \qquad {\rm tr} (i j k l)={\rm tr}_+ (i j k l)+{\rm tr}_- (i j k l)\,.
\ee

In principle, one should be able to directly verify that these one-loop amplitudes satisfy the soft theorem of CS soft theorem eq.~(\ref{CSsoft}), based on the soft expansion of the $H$ function. However, we find it more illuminating to present a proof based on the recursion for all-plus amplitudes, similar to what we have done above for Yang-Mills case. 

We again consider the BCFW representation of the all-plus amplitudes with the holomorphic shift for the soft leg.  Just like the all-plus amplitudes in Yang-Mills, there are two distinct contributions under the BCFW shifts eq.~(\ref{BCFWshift}): the two-particle factorization, and the contribution from the pole at infinity $\mathcal{B}^{\rm G}_n$,\footnote{Recursion relations based on three-leg shifts were used in \cite{Brandhuber:2007up} to eliminate the pole at infinity, but for our purpose we stick with usual two-leg BCFW shifts. }
\eq\label{AllPlusRec}
M^{(1)}_{n+1}(1^+,\cdots, n^+, s^+)= \sum_{1\leq i< n}M^{(0)}_3(\hat{s}^+, i^+,-\hat{K}_{s i}^-)\frac{1}{K^2_{s,i}}M^{(1)}_{n}(\hat{K}_{s i}^+,\cdots,\hat{n}^+)+\mathcal{B}^{\rm G}_n\,.
\eqe 
In the following we will explicitly demonstrate that $\mathcal{B}^{\rm G}_n$ has no soft divergence. Given this, the soft-divergence is solely given by the first term in eq.~(\ref{AllPlusRec}), i.e.:
\eqa
\nonumber M^{(1)}_{n+1}(\epsilon\lambda_s,\tilde{\lambda}_s)\bigg|_{\rm div}&=&\sum_{1\leq i< n}M^{(0)}_3(\hat{s}^+, i^+,-\hat{K}_{s i}^-)\frac{1}{K^2_{s,i}}M^{(1)}_{n}(\hat{K}_{s i}^+,\cdots,\hat{n}^+)\bigg|_{\rm div}\\
\nonumber
%&=&\sum_{1\leq i< n}\frac{1}{\epsilon^3}\frac{\langle ni\rangle^2}{\langle ns\rangle^2}\frac{[si]}{\langle si\rangle}
%e^{ \epsilon\frac{\langle s,n\rangle}{\langle i,n\rangle}\tilde{\lambda}_s\cdot \frac{\partial}{\partial %\tilde{\lambda}_i}  } M^{(1)}_n|_{Div} \\
&=&\left(\frac{1}{\epsilon^3}S_{\rm G}^{(0)}+\frac{1}{\epsilon^2}S_{\rm G}^{(1)}+\frac{1}{\epsilon}S_{\rm G}^{(2)}\right)M^{(1)}_n\,,
\eqae
thus completes the proof of the Cachazo-Strominger soft theorem for all-plus gravity amplitude. We now move on to show that $\mathcal{B}^{\rm G}_n$ is indeed finite in the soft limit. Here we will use the BGK-like representation for $H(a,\{1,\cdots,n\},b)$~\cite{BernAllPlus}:
\eqa
\nonumber H(a,\{1,\cdots,n\},b)&=&\frac{[1,2]}{\langle 1,2 \rangle}\frac{\langle a|K_{1,2}|3]\langle a|K_{1,2,3}|4]\cdots\langle a|K_{1,\ldots,n-1}|n]}{(\langle 23\rangle \cdots \langle n-1n\rangle) (\langle a1\rangle\langle a2\rangle \cdots \langle an\rangle)\langle 1b\rangle\langle nb\rangle}\\
&&+\,\, \mathcal{P}(2,3,\cdots,n)\,,
%\nonumber H(a,\{1,\cdots,n\},b)&=&\frac{[1,2]}{\langle 1,2 \rangle}\frac{\langle a|K_{1,2}|3]\langle a|K_{1,3}|4]\cdots\langle a|K_{1,n-1}|n]}{(\langle 23\rangle \cdots \langle n-1n\rangle) (\langle a1\rangle\langle a2\rangle \cdots \langle an\rangle)\langle 1b\rangle\langle nb\rangle}\\
%&&+\,\, \mathcal{P}(2,3,\cdots,n)\,,
\eqae
%where $K_{1,i}\equiv k_1+k_2+\cdots k_i$, 
and for the special case $n=1$, we have 
\be 
H(a,\{1\},b)= {1 \over (\langle a,1\rangle\langle 1,b\rangle)^2 } \, .
\ee 
Note the potential double poles for $H(a,\{i\},s)$ ($H(a,\{s\},i)$) are cancelled by the corresponding trace factors ${\rm tr}^3(a,\{i\},s,\{i\})$ (${\rm tr}^3(a,\{s\},i,\{s\})$).

Next, we consider the soft behaviour of $\mathcal{B}^{\rm G}_n$. First as we take $z\rightarrow \infty$, since we are shifting $\lambda_s$, any non-trivial soft-divergence must arise from factors of $\langle s\, n\rangle$ in the denominator. Note that although $\langle s \,n\rangle$ appears with degree $-2$ in $H(a,n,s)$ or $H(a,s,n)$, using momentum conservation their corresponding trace factors ${\rm tr}^3(s,\{n\},a,M)$ and ${\rm tr}^3(a,\{s\},n,M)$ is proportional to $\langle s\, n\rangle^3$. One immediately concludes that the boundary term $\mathcal{B}^{\rm G}_n$ can at most have $1/\epsilon$ divergence, and thus establish the validity of both $S_{\rm G}^{(0)}$ and $S_{\rm G}^{(1)}$ for the Cachazo-Strominger soft theorem for all-plus amplitudes in gravity. 

For $S_{\rm G}^{(2)}$ we need to show that all potential $1/\epsilon$ divergences cancel as well. With careful analysis one can conclude that the source for $1/\epsilon$ divergences comes from terms of the following form $H(a,\{M+\hat{n}\},\hat{s})H(a,\{N\},\hat{s}){\rm tr}^3[a,(M+\hat{n}),\hat{s},N]$ and $H(a,\{M+\hat{s}\},\hat{n})H(a,\{N\},\hat{n}){\rm tr}^3[a,(M+\hat{s}),\hat{n},N]$. Let us consider the contributions from each term separately. 
\begin{itemize}
  \item (a): $H(a,\{M+\hat{n}\},\hat{s})H(a,\{N\},\hat{s}){\rm tr}^3[a,(M+\hat{n}),\hat{s},N]$:
   First of all, the trace factor can be written as:
 \eqa
 \nonumber {\rm tr}[a,(M+\hat{n}),\hat{s},N]&=&-{\rm tr}[a,N,\hat{s},N]\;\;\underrightarrow{\quad \lambda_s \rightarrow\epsilon \lambda_s\quad } -z\langle a |N|s]\langle n|N|a]
 \eqae
  Thus the numerator behaves as $z^3$, where as 
  \eqa\label{sub1}
  \nonumber H(a,\{N\},\hat{s}) &=& \frac{H(a,\{N\},n)}{z^2}+\mathcal{O}(z^{-1})\\
  H(a,\{M+\hat{n}\},\hat{s})&=&\frac{\langle a|M|n]}{z\langle ns\rangle\langle an\rangle}H(a,\{M\},n)+\mathcal{O}(z^{0})\,.
  \eqae
  \item (b): $H(a,\{M+\hat{s}\},\hat{n})H(a,\{N\},\hat{n}){\rm tr}^3[a,(M+\hat{s}),\hat{n},N]$:
  This time the trace factor is slightly more complicated:
  \eqa
  \nonumber {\rm tr}[a,(M+\hat{s}),\hat{n},N]=- \langle a |N|n]\langle n|N|a]+z\langle a|N|s]\langle n|N|a]\\
  \eqae
  Thus the numerator behave as $az^3+bz^2+cz+d$, where as:
  \eqa\label{sub2}
  \nonumber H(a,\{N\},n) &=&H(a,\{N\},n)\\
  H(a,\{M+\hat{s}\},\hat{n})&=&\frac{\langle a|M|s]}{z^2\langle sn\rangle\langle an\rangle}H(a,\{M\},n)+\mathcal{O}(z^{-1})\,.
  \eqae
   
\end{itemize}
Note that all subleading terms (in $z$) in eq.(\ref{sub1}) and eq.(\ref{sub2}) do not contribute since at the same time they are subleading in $\epsilon$.\footnote{These terms come from $\l\hat{s}i\r=z\l ni\r(1+\l si\r/z\l ni\r)$, subleading expansion in large $z$ are accompanied by powers of $\l si\r$ which are of order $\epsilon$.} 
Evaluating the order $z^0$ contribution, one finds:
 \eqa
\nonumber (a):&&  -\frac{\langle a|K_M|n]}{\langle ns\rangle\langle an\rangle}H(a,\{M\},n)H(a,\{N\},n)\langle a |K_N|s]^3\langle n|K_N|a]^3 \, ,\\
(b):&&-3\frac{\langle a|K_M|s]}{\langle sn\rangle\langle an\rangle}H(a,\{M\},n)H(a,\{N\},n)\langle a |K_N|n]\langle a|K_N|s]^2\langle n|K_N|a]^3 \, , 
 \eqae
where $K_M$ and $K_N$ denote the sums of the momenta inside the subsets $M$ and $N$, respectively. 
The presence of $1/\langle ns\rangle$ indicates these are of order $\epsilon^{-1}$ at the soft limit. Combining the results $(a)$ and $(b)$, and after simplification we find that soft divergence in $\mathcal{B}^{\rm G}_n$ is given by:
\eqa  
\mathcal{B}^{\rm G}_n|_{\rm div}=
\frac{1}{2}{ [s n]^3 \over \langle s n\rangle  }
\sum_{a,M,N} \langle a n \rangle^2 H(a,\{M\},n)H(a,\{N\},n)\langle a |K_N|n]\langle n|K_N|a]^3 \, .
\eqae
 We have explicitly checked up to $12$ points, the coefficient of the divergent factor $1/\langle s n\rangle$ vanishes under the $n$-point momentum conservation $k_a + K_M + K_N + k_n =0$, namely
 \eqa\label{identity}
 \sum_{a,M,N} \langle a n \rangle^2 H(a,\{M\},n)H(a,\{N\},n)\langle a |K_M|n]\langle n|K_N|a]^3 = 0 \, .
 \eqae
It would be interesting to see if there is a deeper reason for the identity (\ref{identity}) to hold. 

In fact, the $1/\epsilon$ divergence behaves as $1/z^2$ as $z\rightarrow \infty$. To see this let us consider the order $z^{-1}$ term, which only receives contribution from case $(b)$ and is given by:
\eq
3\frac{1}{\langle sn\rangle\langle an\rangle}H(a,\{N\},n)H(a,\{M\},n)\langle a |N|n]^2\langle n|N|a]^3\langle a|N|s]\langle a|M|s]\,.
\eqe
Note that exchanging $M\leftrightarrow N$ the above function obtains a minus sign through momentum conservation. Thus in the full permutation sum, the above term cancel in pair. It is quite interesting that the subleading soft divergence has the same large $z$ asymptotics as the tree-level amplitude.
%%%%%%%%%%%%%%%%%%%%%%%%%%%%%%%%%%%%%%%%%%%%%%%%%
\subsection{Single-minus amplitudes}
%%%%%%%%%%%%%%%%%%%%%%%%%%%%%%%%%%%%%%%%%%%%%%%%%
For single-minus amplitudes in gravity, analytic expressions for up to six-points have been obtained in the literature, by using the so-called ``augmented recursion" involving double-pole contributions~\cite{Dunbar1Minus}. Although the method is similar to the case of single-minus Yang-Mills amplitude, due to the lack of a universal form for the double-pole contribution (the analog of the last line in (\ref{rec_oneminus}) ), it has to be determined on a case-by-case basis, which were explicitly performed for for five- and six-points cases~\cite{Dunbar1Minus}. 

In this note we will not attempt to extensively study the soft graviton theorem for single-minus amplitude, but content ourselves to general argument and explicit checks for five- and six-point cases.  Again there are two soft limits for $M_{n{+}1}(1^-,2^+,\ldots,(n{+}1)^+)$ to consider here: the anti-holomorphic one with $\tilde\lambda_1=\epsilon \tilde\lambda_s$ and the holomorphic one with $\lambda_{n{+}1}=\epsilon \lambda_s$. 

Remarkably, both by (heuristic) analysis based on recursion, and by explicit checks for $n\leq 5$,  we find the soft behavior parallel to the Yang-Mills case: not only $S^{(0)}$ (as expected), but also $S^{(1)}$ receives no quantum corrections in both soft limits, while $S^{(2)}_{\rm G}$, similar to $S^{(1)}_{\rm YM}$) needs modification for the holomorphic soft limit.

Let us first demonstrate the proposal by (heuristic) analysis based on recursions. We consider the shift (\ref{shift2}), and as discussed by Dunbar et al~\cite{Dunbar1Minus}, by assuming no pole at infinity, we have a augmented recursion,
\ba\label{rec_grav_oneminus}
M^{(1)}_{n{+}1} (1^-,2^+,\ldots,(n{+}1)^+)&=&\sum_{1< i< n}M^{(0)}_3((\widehat{n{+}1})^+, i^+,-\hat{K}_{n{+}1,i}^-)\frac{1}{K^2_{n{+}1,i}}M^{(1)}_{n}(\hat{K}_{n{+}1,i}^+,\ldots,\hat{1}^-)\nl
&&+\sum_{L,R} M^{(0)}_L (\hat{1}^-,\{L\}^+,-\hat{K}^-_{1,L}) \frac 1 {K^2_L} M^{(1)}_ R (\hat{K}_{1,L}^+,\{R\}^+,(\widehat{n{+}1})^+)\nl
&&+\sum_{i=1}^n\frac 1 {(K_{i,n{+}1}^2)^2} \left(C^{(i)}_d+K_{i,n{+}1}^2 C^{(i)}_s \right)
\ea 
where the first  two lines represent factorization terms (including the homogenous term), and the third line, for which a universal formula is currently unknown, represents the contributions from double- and simple-pole in $K_{i,n{+}1}^2$, whose coefficients are denoted by $C^{(i)}_d$ and $C^{(i)}_s$. 

Although the explicit forms of $C^{(i)}_d$ and $C^{(i)}_s$ have only been worked out for $n\leq 5$, with some general assumptions, we can still study the soft limits of single-minus amplitude using (\ref{rec_grav_oneminus}). For the soft limit $\tilde\lambda_{1}=\epsilon\tilde\lambda_s \to 0$, completely parallel to the Yang-Mills case, we see that on the first two lines of eq.(\ref{rec_grav_oneminus}), only terms with a single element in $L$ diverges, and it gives the expected soft limit. Therefore, if $C^{(i)}_d$ and $C^{(i)}_s$ have no soft divergences as $\tilde\lambda_{1}\to 0$, we will indeed have the correct soft theorem,
\be
M^{(1)}_{n{+}1}(\{\lambda_s,\epsilon\tilde\lambda_s\}^-,2^+,\ldots,(n{+}1)^+)\bigg|_{\rm div}=\left(\frac{1}{\epsilon^3}\bar{S}_{\rm G}^{(0)}+\frac{1}{\epsilon^2}\bar{S}_{\rm G}^{(1)}+\frac{1}{\epsilon} \bar{S}_{\rm G}^{(2)}\right)M^{(1)}_{n}(2^+,\ldots,(n{+}1)^+),
\ee
where $\bar{S}^{(k)}_{\rm G}$ is  the parity-conjugate of $S^{(k)}_{\rm G}$ for $k=0,1,2$. We have checked explicitly that indeed for $n=4,5$ there are no such soft divergences using the results of~\cite{Dunbar1Minus}, and we assume this continues to be true for all multiplicities. 

On the other hand, for the holomorphic soft limit, $\lambda_{n{+}1}=\epsilon \lambda_s$, we see that the second line has no soft divergences, while the first line alone gives the expected soft theorem. However, similar to the Yang-Mills case, the third line is expected to give $\mathcal{O}(\epsilon^{-1})$ contributions, which is indeed true for $n=4,5$.  Since $K_{i,n{+}1}^2=\epsilon \l i s\r [i n{+}1]$, what we need to assume here is that for any $n$, $C_d$ and $C_s$ behaves as $\mathcal{O}(\epsilon)$ in the limit, and we have explicitly checked this for the five- and six-point results in ~\cite{Dunbar1Minus}. Therefore, based on the two assumptions, we argued that $S^{(0)}$ and $S^{(1)}$ do not receive any quantum corrections for single-minus amplitude, but $S^{(2)}$ does for taking a positive-helicity to be the soft leg. 

We have also directly verified the Cachazo-Strominger soft theorem with all the data available in the literature\footnote{We adopt the normalization of \cite{BernAllPlus}, which differs from that in \cite{Dunbar1Minus} by a factor of $32$.}, namely both soft limits of five- and six-points single-minus amplitude. The four-point amplitude reads~\cite{Bern:1993wt, Dunbar:1994bn} :
\be
M^{(1)}_4(1^-,2^+,3^+,4^+)=\frac 1 {180}\left(\frac {s t }{u }\frac{[2 4]^2}{[1 2]\l 2 3\r [3 4]\l 4 1\r} \right)^2 (s^2 +t^2 +s t)\,.
\ee

We also record here the expression for five-point single-minus amplitude given in~\cite{Dunbar1Minus}. The amplitude is given by the sum of three permutations, $M_5^{(1)}(1^-,2^+,3^+,4^+,5^+)=R(1,2,3,4,5)+R(1,2,4,5,3)+R(1,2,5,3,4)$, with $R=\frac 1 {180} (R_1+R_2+R_3)$ given by three-classes of recursive diagrams of the following form
\ba
R_1(a,b,c,d,e)&=&\frac{\l a d\r^2\l a e\r^2 [b c][d e]^4 \left(\l c d\r^2\l a e\r^2+\l a c\r\l c d\r\l d e\r\l a e\r+\l a c\r^2\l d e\r^2\right)}{\l a b\r^2\l b c\r \l c d\r^2\l c e\r^2\l d e\r^2},\nl
R_2(a,b,c,d,e)&=&-\frac{3 \l a e\r[b e]^4}{\l c d\r^2 [a b]^2 [a e]}\left([b c]^2 [d e]^2+ [b c][c d][d e][b e]+[c d]^2 [b e]^2\right),\nl
R_3(a,b,c,d,e)&=&\frac{\l a b\r^2\l a c\r^4 [b c]^4 [d e]}{\l a d\r\l a e\r\l b c\r^2 \l c d\r\l c e\r\l d e\r} \left(1+ \Delta(a,b,c,d,e)\right),
\ea
where the expression for $\Delta$ can be found in eq. (4.15) of~\cite{Dunbar1Minus}. The expression for six-point single-minus amplitude is given in Appendix A of~\cite{Dunbar1Minus}. 

With these results for single-minus, as well as eq.~(\ref{gra_allplus}) for all-plus amplitudes, we have explicitly checked that Cachazo-Strominger soft graviton theorem holds for all cases except the $S^{(2)}$ contributions for the holomorphic-soft limit of single-minus amplitudes.  

It is desirable to prove both assumptions either by carefully working out the complete recursion, or based on general physical considerations. On the other hand, it would be of great interest to see if the soft theorem can be used to constrain, and together with other constraints such as collinear limits, even fully determine the single-minus amplitudes for gravity.

\section{Discussions \label{sec:Con}}

In this note we studied the new theorem on subleading behavior of soft gravitons~\cite{CS} and gluons~\cite{Casali:2014xpa} at tree and loop level. By reviewing the BCFW proof for trees, some side results we obtained are an infinite series of soft functions for MHV amplitudes, and an interesting double-copy relations between Yang-Mills and gravity soft theorems; we also present generalizations to supersymmetric amplitudes.

We then restrict to infrared-finite, rational amplitudes at one loop, and worked out explicitly subleading terms in the soft expansion of all-plus and single-minus amplitudes in Gravity and Yang-Mills theory. We showed that while the subleading soft behavior of all-plus, as well as single-minus amplitudes with the minus-helicity particle being taken soft, are identical to tree amplitudes, however $S^{(1)}_{\rm YM}$ and $S^{(2)}_{\rm G}$ do require loop corrections for the soft limit of a positive-helicity particle of single-minus amplitudes.

In general loop corrections for the soft behavior are expected for subleading terms, and even for the leading term in Yang-Mills theory, i.e. $S^{(0)}_{\rm YM}$. However, one might expect that such corrections are absent for infared finite observables, such as suitably-defined rational terms for any one-loop amplitudes including those studied here. Our results indicate that, with the exception of $(n{-}1)$-point amplitudes being all-plus, corrections are needed for the subleading soft behavior of rational terms. The simplest example of such corrections may be the $\mathcal{O}(\epsilon^{-1})$-mismatch for holomorphic soft limit of single-minus amplitudes, eq.~(\ref{corr}) for Yang-Mills and the corresponding mismatch for five- and six-graviton cases. 

We have traced the origin of such mismatch to the new contributions containing double-poles of two-particle factorization, first appearing for one-loop single-minus amplitudes. Since such contributions are ubiquitous for recursions of rational terms in gauge theories~\cite{Bern1Minus}, we strongly suspect that corrections for $S^{(1)}_{\rm YM}$ is needed there as well. Moreover, corrections for $S^{(2)}_{\rm G}$ are also expected for rational terms of one-loop amplitudes in supergravity theories, in particular the remarkable formula for that in $\mathcal{N}=4$ supergravity~\cite{Dunbar:2011dw, Dunbar:2012aj}. We leave the study of such corrections, and their possible physical interpretations, to the future.

Very recently in~\cite{Bern:2014oka}, leading infrared-singular loop corrections to the subleading soft behavior have been obtained. It is desirable to combine their results with our infrared-finite corrections, to understand the complete subleading soft behavior at loop level. In addition, the new soft theorem is claimed to be a consequence of the BMS symmetry. With a better understanding of its loop corrections, it would be very interesting to investigate the status of the symmetry at quantum level. The possible connection to Yangian generator for $\mathcal{N}=4$ super Yang-Mills is also intriguing, and may deserve further study.

As we have mentioned, the new soft theorem can be extended to a wider context, e.g. gauge theories and gravity in any dimensions, as well as supersymmetric theories. For the latter, a particularly interesting application is to the soft scalar emission of $\mathcal{N}=8$ supergravity, at both tree and loop level. It is well known that the emission of single and multiple soft scalars is related to the $E_{7(7)}$ symmetry~\cite{ArkaniHamed:2008gz}, which is expected to hold at quantum level. It would be of great interest to study single and multiple soft scalar behavior, including subleading terms, and possible relations to $E_{7(7)}$ symmetry (e.g. the single-soft case at one-loop was studied in \cite{He:2008pb}). 

In~\cite{Schwab:2014xua}, it was argued that soft theorems in arbitrary dimensions follow from a new representation of tree amplitudes in terms of scattering equations~\cite{CHY}, and we point out that the double-copy relation between soft gluon theorem and soft graviton theorem is also manifest in that representation. On the other hand, it is very tempting to use recursion relations in arbitrary dimensions to prove the new theorem, and we take a first step in this direction here. 

\subsection{Higher-dimensional soft theorems}

An interesting aspect in the analytic proof of tree-level soft theorems, is the fact that subleading corrections $S_{\rm G}^{(1,2)}$ and $S_{\rm YM}^{(1)}$ are isolated to the two-particle factorization diagrams including the soft leg. An interesting question is whether this continues to be true in higher dimensions. A suitable form of BCFW recursion for general dimensions was presented by Arkani-Hamed and Kaplan~\cite{ArkaniHamed:2008yf}, where the momenta is shifted by a light-like vector that is associated with the polarization vectors:
\eqa
\nonumber k_{\hat{s}}=k_s+qz,\quad k_{\hat{n}}=k_n-qz,\quad  \epsilon_s^-=\epsilon_n^+=q,\\
\;\; \epsilon_s^+=q^*-zk_n,\;\epsilon_n^-=q^*+zk_s,\quad \epsilon_T=(0,0,\cdots,1,\cdots,0)
\eqae
where we are in the Loretnz frame such that leg $s$ and $n$ are back to back, and $\epsilon_T$ are the $ D{-}4$ distinct basis for the transverse directions. Let us consider the soft leg being $\epsilon_s^-$, where BCFW recursion is valid for any polarization state that leg-$n$ belongs to. The resulting Yang-Mills amplitude is then given by 
\eqa
\nonumber A_{n+1}&=&\sum_{\rm states \,I}\sum_{1 {\leq} i<n{-}1}A_{p}(\{k_{\hat{s}},\epsilon_s\}\cdots,,\{\epsilon_i,k_i\}, I)\frac{1}{K^2_{s,1,\cdots.i}}A_{m}(-I,\cdots, \{k_{\hat{n}}, \hat{\epsilon}_n\})\\
&=&\sum_{\rm pol}\sum_{1 \leq i<n-1}\frac{(\epsilon_{s\mu}a^{\mu\nu}_{p}\epsilon_{I\nu})(\epsilon_{I\rho}a_{{m}}^{\rho\sigma}\hat{\epsilon}_{n\sigma})}{K^2_{s,1,\cdots.i}}
\eqae
where $p+m=n+3$ and $\sum_{\rm pol}$ represents that we sum over polarization states for $\epsilon_I$. On the other hand, from the Ward identity on the leg $s$ in $A_{p}$ we can deduce:
\eqa
k_{\hat{s}\mu}a^{\mu\nu}_{p}\epsilon_{I\nu}=0 \rightarrow q_\mu a^{\mu\nu}_{p}\epsilon_{I\nu}= -\frac{1}{z}k_{s\mu}a^{\mu\nu}_{p}\epsilon_{I\nu}\,.
\eqae
Now since $\epsilon_s^-=q$, from the above we immediately see that $\epsilon_{s\mu}^- a^{\mu\nu}_{p}\epsilon_{I\nu}$ is in fact proportional to the soft momenta $k_s$ and thus vanishes at order $\epsilon$ in the soft limit \textit{except} for when the factorization channel are two particle poles containing the soft particle.\footnote{There will still be a soft divergence since the solution for $z$ in this channel is also of order $\epsilon$.} The same analysis can be applied to gravity. Higher dimensional soft theorems are interesting in itself, as it implies non-trivial soft theorems for massive amplitudes in four-dimensions. We leave the exploration on this front to future work.  
%%%%%%%%%%%%%%%%%%%%%%%%%%%%%%%%%%%%%%%%%%%%%%%%%%%%%
\section{Acknowledgements}
%%%%%%%%%%%%%%%%%%%%%%%%%%%%%%%%%%%%%%%%%%%%%%%%%%%%%%
It is a pleasure to thank Marcus Spradlin and  Anastasia Volovich for private communication, and Andreas Brandhuber and Gabriele Travaglini for helpful discussion. S. H thanks Tristan McLoughlin for useful discussions. The work of S. H is supported by Zurich Financial Services Membership and the Ambrose Monell Foundation. The work of Y-t. H is supported by the National Science Foundation Grant PHY-1314311 . The work of C.W is supported by the Science and Technology Facilities Council Consolidated Grant ST/J000469/1 {\it String theory, gauge theory \& duality. }

\end{document}